\documentclass[conference]{IEEEtran}
\IEEEoverridecommandlockouts
\usepackage{cite}
\usepackage{amsmath,amssymb,amsfonts}
\usepackage{algorithmic}
\usepackage{graphicx}
\usepackage{textcomp}
\usepackage{xcolor,colortbl}
\usepackage{scalerel}
\usepackage{graphics}
\usepackage{academicons}
\usepackage{tikz}
\usepackage[utf8]{inputenc}
\usepackage[hidelinks]{hyperref}
\usepackage[flushleft]{threeparttable}
\usepackage{algorithm}
\usepackage{array}
\usepackage{multirow}
\usepackage{booktabs}
\usepackage{color}
\usepackage{array}
\usepackage{lipsum}
\usepackage{nccmath}
\usetikzlibrary{shapes,arrows}
\usepackage{eso-pic}
\usepackage{pdfpages}

\definecolor{LightCyan}{rgb}{0.88,1,1}
\usetikzlibrary{svg.path}

\definecolor{orcidlogocol}{HTML}{A6CE39}
\tikzset{
    orcidlogo/.pic={
        \fill[orcidlogocol] svg{M256,128c0,70.7-57.3,128-128,128C57.3,256,0,198.7,0,128C0,57.3,57.3,0,128,0C198.7,0,256,57.3,256,128z};
        \fill[white] svg{M86.3,186.2H70.9V79.1h15.4v48.4V186.2z}
        svg{M108.9,79.1h41.6c39.6,0,57,28.3,57,53.6c0,27.5-21.5,53.6-56.8,53.6h-41.8V79.1z M124.3,172.4h24.5c34.9,0,42.9-26.5,42.9-39.7c0-21.5-13.7-39.7-43.7-39.7h-23.7V172.4z}
        svg{M88.7,56.8c0,5.5-4.5,10.1-10.1,10.1c-5.6,0-10.1-4.6-10.1-10.1c0-5.6,4.5-10.1,10.1-10.1C84.2,46.7,88.7,51.3,88.7,56.8z};
    }
}

\newcommand\orcidicon[1]{\href{https://orcid.org/#1}{\mbox{\scalerel*{
                \begin{tikzpicture}[yscale=-1,transform shape]
                \pic{orcidlogo};
                \end{tikzpicture}
            }{|}}}}

\def\BibTeX{{\rm B\kern-.05em{\sc i\kern-.025em b}\kern-.08em
    T\kern-.1667em\lower.7ex\hbox{E}\kern-.125emX}}

\IEEEoverridecommandlockouts
\IEEEpubid{\makebox[\columnwidth]{\textbf{978-1-5386-5541-2/18/\$31.00~\copyright2021 IEEE} \hfill} \hspace{\columnsep}\makebox[\columnwidth]{ }}
 
\begin{document}

\begin{titlepage}
\includepdf{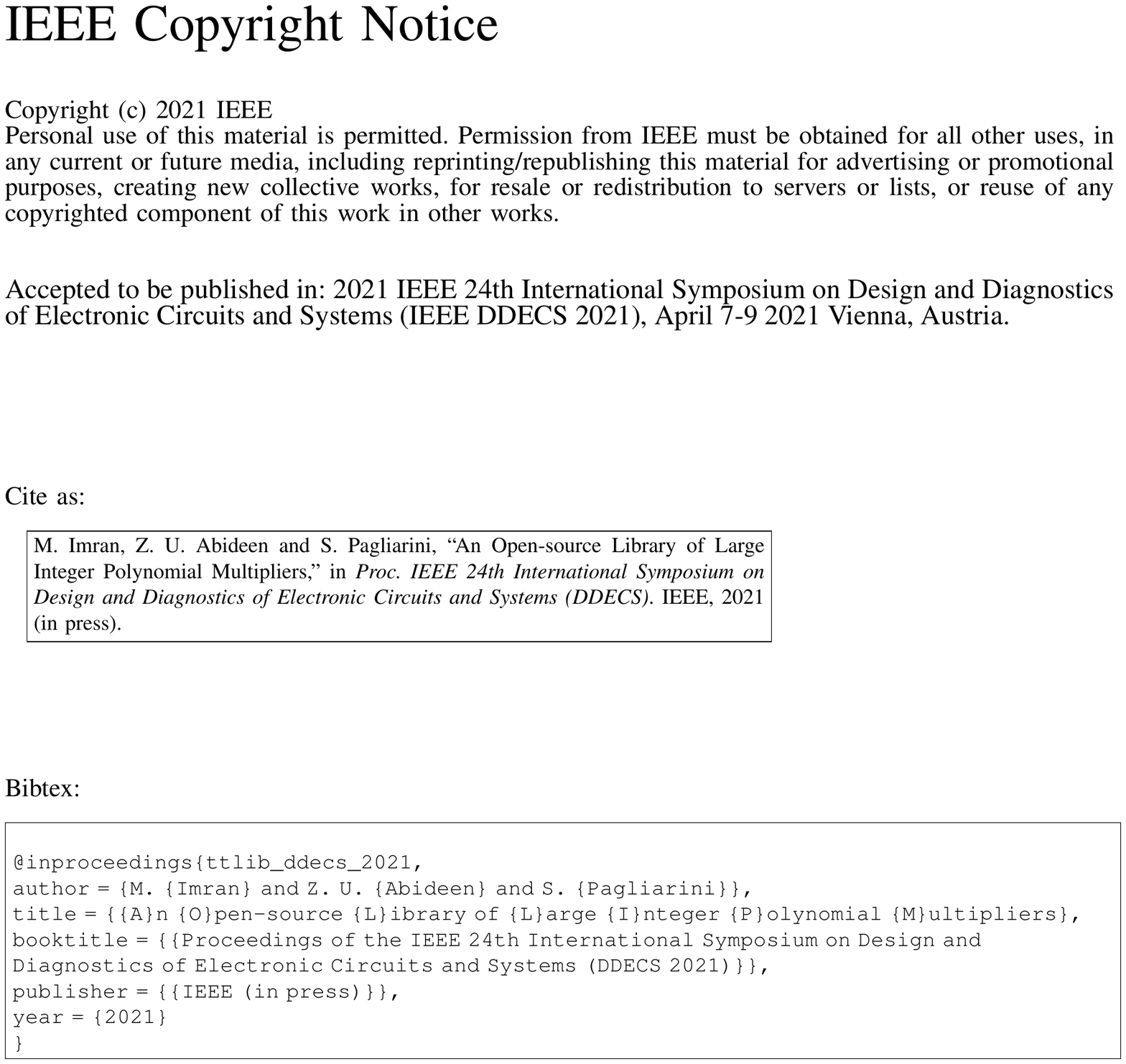}
\end{titlepage}

\title{An Open-source Library of Large Integer Polynomial Multipliers
\thanks{This work was partially supported by the EC through the European Social Fund in the context of the project ``ICT programme". It was also partially supported by the Estonian Research Council grant MOBERC35.}
}

\author{\IEEEauthorblockN{Malik Imran}
\IEEEauthorblockA{\textit{Centre for Hardware Security} \\
\textit{Tallinn University of Technology}\\
Tallinn, Estonia \\
malik.imran@taltech.ee}
\and
\IEEEauthorblockN{Zain Ul Abideen}
\IEEEauthorblockA{\textit{Centre for Hardware Security} \\
\textit{Tallinn University of Technology}\\
Tallinn, Estonia \\
zain.abideen@taltech.ee}
\and
\IEEEauthorblockN{Samuel Pagliarini}
\IEEEauthorblockA{\textit{Centre for Hardware Security} \\
\textit{Tallinn University of Technology}\\
Tallinn, Estonia \\
samuel.pagliarini@taltech.ee}
}

\maketitle
\IEEEpubidadjcol
\begin{abstract}
Polynomial multiplication is a bottleneck in most of the public-key cryptography protocols, including Elliptic-curve cryptography and several of the post-quantum cryptography algorithms presently being studied. In this paper, we present a library of various large integer polynomial multipliers to be used in hardware cryptocores. Our library contains both digitized and non-digitized multiplier flavours for circuit designers to choose from. The library is supported by a C++ generator that automatically produces the multipliers' logic in Verilog HDL that is amenable for FPGA and ASIC designs. Moreover, for ASICs, it also generates configurable and parameterizable synthesis scripts. The features of the generator allow for a quick generation and assessment of several architectures at the same time, thus allowing a designer to easily explore the (complex) optimization search space of polynomial multiplication.
\end{abstract}


\section{Introduction}
Polynomial multiplication (i.e., $c(x)=a(x)\times b(x)$) is a fundamental building block for cryptographic hardware and is often identified as the bottleneck in implementing efficient circuits. The most widely deployed public key crypto systems (e.g., RSA and ECC) need polynomial multiplications \cite{RSA_ECC}. Many of the post-quantum cryptography (PQC) algorithms (e.g., NTRU-Prime, FrodoKEM, Saber, etc.) also require large integer multipliers for multiplying polynomial coefficients utilized to perform key-encapsulations and digital signatures \cite{NIST_Competition}. Another application is in fully homomorphic encryption, a specific branch of cryptography that requires large integer multipliers to enable multi-party and secure-by-construction on the cloud computations \cite{cloud-computations}. There is a clear demand for large integer multipliers to perform multiplication over polynomial coefficients. \textbf{To our knowledge, today, no widely available repository of open source multiplier architectures exists}. This is the gap that our library addresses. 

There are several multiplication methods employed to perform multiplication over polynomial coefficients, including the schoolbook method (SBM), Karatsuba, Toom-Cook, Montgomery, and number theoretic transformation (NTT). A quick scan of the PQC algorithms involved in the NIST standardization effort \cite{NIST_Competition_Round_2} reveals that many reference implementations suggest the use of these multipliers: SBM is suggested by the authors of NTRU-Prime and FrodoKEM, Karatsuba and Toom-Cook methods are used in Saber and NTRU, a combination of NTT and SBM is suggested for CRYSTALS-Kyber, SBM and Montgomery are considered in Falcon. 

Examples of recent works employing non-digitized and digitized polynomial multiplication methods are given in \cite{Rafferty_2017,2_way_Karatsuba,Montgomery,NTT,ASIC_65nm_2014,Interleaved_modular_reduction_algorithm,homomorphic_encryption} and \cite{Xie_2018,Morales_Sandoval,Pan}, respectively. In \cite{Rafferty_2017}, for different polynomial sizes, an architectural evaluation of different multiplication methods (SBM, comba, Karatsuba, Toom-Cook, Montgomery, and NTT) is performed over a Virtex-7 FPGA platform. An improved Montgomery polynomial multiplier is presented in \cite{Montgomery} for a polynomial size of 1024 bits over a Virtex-6 FPGA. A run-time configurable and highly parallelized NTT-based polynomial multiplication architecture over Virtex-7 is discussed in \cite{NTT}. A systolic based digit serial multiplier wrapper on an Intel Altera Stratix-V FPGA is described in \cite{Xie_2018}, where digit sizes of 22 and 30 bits are considered for operand lengths 233 and 409 bits, respectively. A digit serial Montgomery based wrapper is provided in \cite{Morales_Sandoval}, where a digit size of 64 is selected for the operand length 571 bits, on a Virtex-6. Similarly, a digit serial modular multiplication based wrapper on Virtex-7 is shown in \cite{Pan}, where digit sizes of 2, 4 and 8 bits are preferred for an operand length of 2048 bits. 

ASIC implementations, while less frequent, also explore the polynomial multiplication design space. In \cite{2_way_Karatsuba}, different polynomial multipliers with different operand lengths are considered for area and power evaluations on a 65nm technology. On similar technology, a bit level parallel-in-parallel-out (BL-PIPO) multiplier architecture and a modified interleaved modular reduction multiplication with bit-serial sequential architecture is proposed in \cite{Interleaved_modular_reduction_algorithm, ASIC_65nm_2014}, respectively. Using a 65nm commercial node,  for an operand length of 409 bits. For fully homomorphic encryption schemes, an optimized multi-million bit multiplier
based on the Schonhage Strassen multiplication algorithm is described in \cite{homomorphic_encryption} on 60nm technology node.

Although there are several reported implementations of different multiplication methods \cite{Rafferty_2017,2_way_Karatsuba,Montgomery,NTT,ASIC_65nm_2014,Interleaved_modular_reduction_algorithm,homomorphic_encryption,Xie_2018,Morales_Sandoval,Pan}, these implementations tend to be specifically tailored for a given operand size and for a given target (e.g., high speed or low area). The matter is that this trade-off space is rather complicated to navigate without automation. Consequently, a common approach to assess (several) multiplication methods is required.

In order to tackle the aforementioned limitations of the available literature and the need for automation, we develop an open-source library of multipliers which we name TTech-LIB. Our library is supported by a C++ generator utility that produces -- following user specifications -- hardware description of four selected multiplication methods: (a) SBM, (b) 2-way Karatsuba, (c) 3-way Toom-Cook, and (d) 4-way Toom-Cook. For selected multiplication methods, our library also offers a digitized solution: a single parameterized digit-serial wrapper to multiply polynomial coefficients. By default, the wrapper instantiates a singular SBM multiplier, but it can be replaced by any other multiplier method since the interfaces are identical between all methods. Finally, FPGA and ASIC designers can select their own multiplication method, size of the input operands, and digit size (only for the digitized wrapper, naturally). Moreover, for ASIC designers, there is the possibility to generate synthesis scripts for one of two synthesis tools, either Cadence Genus or Synopsys Design Compiler (DC). The user is not restricted to generating a single architecture at a time, the generator will produce multiple solutions if asked to do so, which will appear as separate Verilog (.v) files.

The remainder of this work is structured as follows: The mathematical background for selected multiplication methods is described in Section \ref{mathematical_background}. The generator architecture and the structure of proposed TTech-LIB is provided in Section \ref{TTech_LIB}. Section \ref{sec:results} shows the experimental results and provide comparisons of non-digitized and digitized flavours of multiplication methods. Finally, Section \ref{conclusion} concludes the paper.

\section{Mathematical background} \label{mathematical_background}
In this section, we present the mathematical formulations behind polynomial multiplication. We assume the inputs are two $m$-bit polynomials and the output is a polynomial of size $2m-1$.

\subsection{Non-digitized multiplication} The SBM is the traditional way to multiply two input polynomials $a(x)\times b(x)$, as shown in Eq. \ref{eq:eq_sbm}. To produce resultant polynomial $c(x)$ by performing bit by bit operations, it requires $2\times m$ clock cycles, $m^2$ multiplications and $(m-1)^2$ additions.
    
    \begin{figure}[ht]
        \begin{ceqn}
        \begin{align}\label{eq:eq_sbm}
            c(x)=\sum_{i=0}^{m-1}\sum_{j=0}^{m-1}a\textsubscript{i}b\textsubscript{j}x^{i+j}
        \end{align}
        \end{ceqn}
    \end{figure}
    
Other approaches such as the 2-way Karatsuba, 3-way Toom-Cook, and 4-way Toom-Cook are more time efficient since they split the polynomials into $n$ equal parts, as shown in Eq. \ref{eq:eq_2_way_KM}. The value of $n$ for 2-way Karatsuba, 3-way Toom-Cook and 4-way Toom-Cook multipliers is 2, 3 and 4, respectively and as the name implies. In Eq. \ref{eq:eq_2_way_KM}, the variable $k$ determines the index of the split input polynomial. For example, for a 4-way Toom-Cook multiplier, the values of $k$ are \{3, 2, 1, 0\}, meaning the input polynomial $a(x)$ becomes $a_3(x)$, $a_2(x)$, $a_1(x)$, and $a_0(x)$.

    \begin{figure}[ht]
        \begin{ceqn}
        \begin{align}\label{eq:eq_2_way_KM}
            \resizebox{0.445\textwidth}{!}{$c(x)=\underbrace{\left(\sum_{i={\frac{k \times m}{n}}}^{m-1} a\textsubscript{k}(x) + \ldots + \sum_{i=0}^{\frac{k \times m}{n}-1} a\textsubscript{0}(x)\right)}_\text{$split\, polynomial\,  a(x)$}
            \times
            \underbrace{\left(\sum_{i={\frac{k \times m}{n}}}^{m-1} b\textsubscript{k}(x) + \ldots + \sum_{i=0}^{\frac{k \times m}{n}-1} b\textsubscript{0}(x)\right)}_\text{$split\, polynomial\, b(x)$}$}
        \end{align}
        \end{ceqn}
    \end{figure}
    
In Eq. \ref{eq:eq_2_way_KM_1}, the expanded version of Eq. \ref{eq:eq_2_way_KM} is presented for the case of 2-way split of input polynomials. The straightforward computation would require four multiplications: (1) one for the computation of inner product resulting polynomial $c_1(x)$, two multiplications for the computation of $c_2(x)$, and finally one multiplication for the computation of $c_0(x)$. However, $c_2(x)$ could be alternatively calculated with only one multiplication, as shown in Eq. \ref{eq:eq_2_way_KM_2}. This is the Karatsuba observation. To generate the final resultant polynomial $c(x)$, addition of inner products is required, as presented in Eq. \ref{eq:eq_2_way_KM_3}. Similarly, when considering the 3-way and 4-way Toom-Cook multipliers, the expanded versions of Eq. \ref{eq:eq_2_way_KM} produce nine and sixteen multiplications, respectively. These multiplications are then reduced to five and seven using a process similar to the 2-way Karatsuba, respectively. We omit the equations for Toom-Cook multipliers for the sake of brevity.


    \begin{figure}[ht]
        \begin{ceqn}
        \begin{align}\label{eq:eq_2_way_KM_1}
            \resizebox{0.40\textwidth}{!}{$c(x)= \underbrace{a\textsubscript{1}(x) b\textsubscript{1}(x)}_{\color{black}\text{$c_1(x)$}} + \underbrace{a\textsubscript{1}(x) b\textsubscript{0}(x) + a\textsubscript{0}(x) b\textsubscript{1}(x)}_{\color{black}\text{$c_2(x)$}} + 
            \underbrace{a\textsubscript{0}(x) b\textsubscript{0}(x)}_{\color{black}\text{$c_0(x)$}}$}
        \end{align}
        \end{ceqn}
    \end{figure}
    
    \begin{figure}[ht]
        \begin{ceqn}
        \begin{align}\label{eq:eq_2_way_KM_2}
            \resizebox{0.42\textwidth}{!}{$c\textsubscript{2}(x) = (a\textsubscript{1}(x) + a\textsubscript{0}(x)) \times (b\textsubscript{1}(x) + b\textsubscript{0}(x)) -  c\textsubscript{1}(x) - c\textsubscript{0}(x)$}
        \end{align}
        \end{ceqn}
    \end{figure}
    
    
    \begin{figure}[ht]
        \begin{ceqn}
        \begin{align}\label{eq:eq_2_way_KM_3}
            \resizebox{0.22\textwidth}{!}{$c(x)= c\textsubscript{0}(x) + c\textsubscript{1}(x) + c\textsubscript{2}(x)$}
        \end{align}
        \end{ceqn}
    \end{figure}
    
    Now, let us assume that the polynomials involved in the multiplications above remain relatively large in size even after split. Thus, SBM multipliers can be employed to resolve the partial products. For a 2-way Karatsuba multiplier of $m$-bit input polynomials, there will be 3 SBM multipliers and each will take two polynomials of size $\frac{m}{2}$ as inputs. Each multiplier requires $\frac{m}{2}$ clock cycles to be completed. If all multipliers operate in parallel, the overall computation also takes $\frac{m}{2}$ cycles. For 3-way and 4-way splits, the number of clock cycles is $\frac{m}{3}$ and $\frac{m}{4}$, respectively. Since our library is aimed at large polynomials, the 2-way Karatsuba, 3-way Toom-Cook, and 4-way Toom-Cook codes available in it actually implement the parallel SBM strategy discussed above. In fact, our non-digitized multipliers are \textbf{hybrid} multipliers.
    
\subsection{Digitized multiplication} The digit serial wrapper in TTech-LIB takes two $m$-bit polynomials $a(x)$ and $b(x)$ as an input and produces $c(x)$ as an output. Digits are created for polynomial $b(x)$ with different sizes which are user-defined as follows: $d=\frac{m}{n}$, where $d$ determines the total number of digits, $m$ denotes the size of input polynomial $b(x)$, and $n$ is the size of each digit. Then, the multiplication of each created digit is performed serially with the input polynomial $a(x)$, while the final resultant polynomial $c(x)$ is produced using shift and add operations. The main difference here is that our digitized solution is serial, while the 2-, 3-, and 4-way multipliers are parallel. The required computational cost (in clock cycles) to perform one digit multiplication is $n$. Since there are $d$ digits, the overall computation takes $d\times n$ clock cycles. It is important to mention that users/designers can choose any multiplication method inside the described digit serial wrapper as per their application requirements. We have used an SBM multiplication method as default.

\section{How to access TTech-LIB} \label{TTech_LIB}
\begin{figure}[]
\centering \footnotesize
\includegraphics[width=2.4in,height=2.6in]{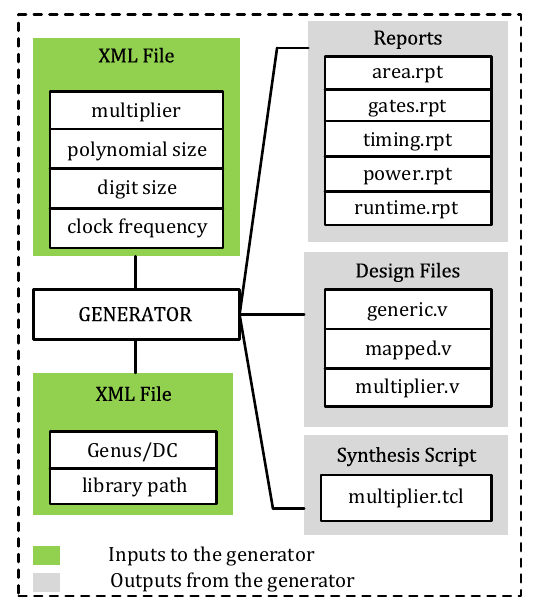}\caption{Generator architecture and file structure of TTech-LIB}
\centering
\label{fig:figure_1}
\end{figure}

    The complete project files (written in C++) are freely available to everyone on our GitHub repository \cite{TTech_LIB}. A sample of pre-generated multipliers is also included in the repository. As shown in Fig. \ref{fig:figure_1}, the user settings can be customized by using a configuration file (\textit{config.xml}). The structure of the library is rather simple and includes five  directories: (1) bin, (2) run, (3) src, (4) synth, and (5) vlog. After running the generator binary, the produced synthesis scripts are put in the synth directory while the generated multipliers are put in the vlog folder. All generated multipliers have the same interface (i.e., inputs are $clk$, $rst$, $a$, and $b$; the output is $c$).



\section{Experimental Results and Comparisons} \label{sec:results}
\subsection{Implementation results and evaluations}
The experimental results for non-digitized and digitized polynomial multiplication methods over NIST defined field lengths \cite{NIST_ECC_PARAMETERS} on 65nm technology node using Genus, Cadence is provided in Table \ref{tab:table_1} and Table \ref{tab:table_2}, respectively. Moreover, the implementation results for various digit sizes of digitized SBM multiplication method over an Artix-7 FPGA device is given in Table \ref{tab:table_3}. In tables \ref{tab:table_1}--\ref{tab:table_2}, clock frequency (\textit{MHz}), area (in $\mu m^2$), and power (\textit{mW}) values are achieved after synthesis using Cadence Genus. Similarly, in Table \ref{tab:table_3}, clock frequency (\textit{MHz}), look-up-tables (LUTs), utilized registers (Regs) and power (\textit{mW}) values are achieved after synthesis using Vivado design tool. Finally, latency for both digitized and non-digitized multipliers (in tables \ref{tab:table_1}--\ref{tab:table_3}) is calculated using Eq. \ref{eq:latency}:

    
    \begin{figure}[ht]
        \begin{ceqn}
        \begin{align}\label{eq:latency}
            \resizebox{0.38\textwidth}{!}{$latency\,(\mu s) = \underbrace{\underbrace{\frac{clock\,cycles}{frequency\,(MHz)}}_{\color{blue}\text{non-digitized}} \times \,total\,digits}_{\color{blue}\text{digitized}}$}
        \end{align}
        \end{ceqn}
    \end{figure}
\begin{table}[ht]
\begin{threeparttable}
\caption{Results of non-digitized multipliers for NIST recommended Elliptic curves over prime and binary fields} \label{tab:table_1}
\begin{tabular}{|p{2.5cm}|p{0.8cm}|p{0.6cm}|p{0.7cm}|p{1.0cm}|p{0.7cm}|}
\hline
\textit{\textbf{Multiplier}} & \textit{\textbf{m}} & \textit{\textbf{Freq (MHz)}} & \textit{\textbf{latency ($\mu s$)}} & \textit{\textbf{Area ($\mu m^2$)}} & \textit{\textbf{Power (mW)}}\\\hline
\multirow{10}{*}{\textbf{Schoolbook}} & P-192 & 500 & {0.382} & 32011.2 & 13.8 \\ \cline{2-6}
{} & {P-224} & 486 & {0.458} & 38048.0 & 17.1 \\ \cline{2-6}
{} & {P-256} & 480 & {0.531} &  48726.7 & 16.9 \\ \cline{2-6}
{} & {P-384} & 444 & {0.862} &  67861.8 & 27.1 \\ \cline{2-6}
{} & {P-521} & 434 & {1.198} &  100242.0 & 28.0 \\ \cline{2-6}
{} & {B-163} & 500 & {0.324} &  29341.4 & 12.9 \\ \cline{2-6}
{} & {B-233} & 476 & {0.487} &  39321.4 & 16.0 \\ \cline{2-6}
{} & {B-283} & 454 & {0.621} &  50603.4 & 17.8 \\ \cline{2-6}
{} & {B-409} & 442 & {0.923} &  73587.6 & 28.2 \\ \cline{2-6}
{} & {B-571} & 413 & {1.380} &  89993.2 & 29.1 \\ \cline{1-6}

\multirow{10}{*}{\textbf{2-way Karatsuba}} & P-192 & 473 & {0.202} & 41379.5 & 8.2 \\ \cline{2-6}
{} & {P-224} & 469 & {0.238} & 49514.4 & 9.6 \\ \cline{2-6}
{} & {P-256} & 467 & {0.274} &  59532.1  & 11.8 \\ \cline{2-6}
{} & {P-384} & 420 & {0.457} & 74844.0  & 15.2 \\ \cline{2-6}
{} & {P-521} & 408 & {0.639} & 105059.5  & 20.8 \\ \cline{2-6}
{} & {B-163} & 487 & {0.168} & 35060.0  & 7.7 \\ \cline{2-6}
{} & {B-233} & 478 & {0.244} & 52328.2 & 10.0 \\ \cline{2-6}
{} & {B-283} & 455 & {0.312} & 64743.8  & 12.6 \\ \cline{2-6}
{} & {B-409} & 432 & {0.474} & 84778.6 & 17.2 \\ \cline{2-6}
{} & {B-571} & 418 & {0.684} & 120374.3 & 21.7 \\ \cline{1-6}

\multirow{10}{*}{\textbf{3-way Toom-Cook}} & P-192 & 909 & {0.070} & 96498.4 & 44.4 \\ \cline{2-6}
{} & {P-224} & 869 & {0.086} & 102470.8 & 46.9 \\ \cline{2-6}
{} & {P-256} & 826 & {0.104} & 104820.9 & 49.4  \\ \cline{2-6}
{} & {P-384} & 689 & {0.185} & 139375.1  & 57.2 \\ \cline{2-6}
{} & {P-521} & 680 & {0.255} & 201341.2  & 80.0 \\ \cline{2-6}
{} & {B-163} & 934 & {0.058} & 75085.6  & 36.0 \\ \cline{2-6}
{} & {B-233} & 877 & {0.088} & 106357.7  & 49.5 \\ \cline{2-6}
{} & {B-283} & 800 & {0.118} &  115188.1 & 54.5 \\ \cline{2-6}
{} & {B-409} & 775 & {0.176} & 170509.0  & 78.4 \\ \cline{2-6}
{} & {B-571} & 766 & {0.249} &  256604.4 & 115.9 \\ \cline{1-6}

\multirow{10}{*}{\textbf{4-way Toom-Cook}} & P-192 & 900 & {0.053} & 105679.1 & 56.9 \\ \cline{2-6}
{} & {P-224} & 847 & {0.066} & 125124.1 & 62.0  \\ \cline{2-6}
{} & {P-256} & 826 & {0.077} & 122298.1 & 63.6 \\ \cline{2-6}
{} & {P-384} & 793 & {0.121} & 241893.7 & 98.2 \\ \cline{2-6}
{} & {P-521} & 767 & {0.170} & 332534.9 & 139.4 \\ \cline{2-6}
{} & {B-163} & 925 & {0.044} & 94834.1  & 49.9 \\ \cline{2-6}
{} & {B-233} & 892 & {0.066} & 132080.0 & 64.2 \\ \cline{2-6}
{} & {B-283} & 826 & {0.085} & 145709.3 & 70.6 \\ \cline{2-6}
{} & {B-409} & 769 & {0.133} & 236989.4 & 99.0 \\ \cline{2-6}
{} & {B-571} & 746 & {0.191} & 340750.8 & 148.2 \\ \cline{1-6}
\end{tabular}
\begin{tablenotes}
      \small
      \item \textit{m} determines the field size or length of the inputs (in bits), where `P' stands for Prime and `B' stands for Binary
    \end{tablenotes}
  \end{threeparttable}
\end{table}

\begin{table}[ht]
\caption{Results of digitized multipliers for NIST recommended Elliptic curves over prime and binary fields} \label{tab:table_2}
{\begin{tabular}{|p{0.2cm}|p{0.8cm}|p{0.9cm}|p{0.7cm}|p{0.7cm}|p{1.0cm}|p{0.7cm}|}\hline
\textit{\textbf{m}} & \textit{\textbf{digit size (n)}} & \textit{\textbf{total digits (d)}} & \textit{\textbf{Freq (MHz)}} & \textit{\textbf{latency ($\mu s$)}} & \textit{\textbf{Area ($\mu m^2$)}} & \textit{\textbf{Power (mW)}}\\\hline

\multirow{4}{*}{\rotatebox[origin=c]{90}{521$\times$521}} & 32 & 17 & 505 & 1.07 & 106956.7 & 30.9 \\ \cline{2-7}
{} & 41 & 13 & 377 & 1.41 & 101538.7 & 26.1 \\ \cline{2-7}
{} & 53 & 10 & 340 & 1.55 & 94752.7 & 20.0 \\ \cline{2-7}
{} & 81 & 7 & 336 & 1.68 & 84321.0 & 15.4 \\ \hline

\multirow{4}{*}{\rotatebox[origin=c]{90}{571$\times$571}} & 32 & 18 & 487 & 1.18 & 114999.8 & 36.7 \\ \cline{2-7}
{} & 41 & 14 & 369 & 1.55 & 116010.3 & 28.9 \\ \cline{2-7}
{} & 53 & 11 & 312 & 1.86 & 91393.9 & 18.1 \\ \cline{2-7}
{} & 81 & 8 & 291 & 2.22 & 76146.8 & 14.1 \\ \hline

\multirow{10}{*}{\rotatebox[origin=c]{90}{1024$\times$1024}} & 2 & 512 & 363 & 2.82 & 196131.2 & 38.0 \\ \cline{2-7}
{} & 4 & 256 & 357 & 2.86 & 178581.2 & 35.1 \\ \cline{2-7}
{} & 8 & 128 & 353 & 2.90 & 167536.4 & 31.5 \\ \cline{2-7}
{} & 16 & 64 & 343 & 2.98 & 166533.1 & 30.2 \\ \cline{2-7}
{} & 32 & 32 & 313 & 3.27 & 148489.5 & 23.0 \\ \cline{2-7}
{} & 64 & 16 & 285 & 3.59 & 122257.8 & 20.8 \\ \cline{2-7}
{} & 128 & 8 & 268 & 3.82 & 123164.6 & 19.9 \\ \cline{2-7}
{} & 256 & 4 & 263 & 3.89 & 129542.4 & 19.5 \\ \cline{2-7}
{} & 512 & 2 & 261 & 3.92 & 136292.4 & 23.1 \\ \cline{2-7}
{} & 1024 & 1 & 259 & 3.95 & 177834.2 & 24.1 \\ \hline

\end{tabular}}
\end{table}

\begin{table}[ht]
\caption{FPGA based results of digitized 1024$\times$1024 SBM multiplier for different digit sizes (Artix-7)} \label{tab:table_3}
{\begin{tabular}{|p{0.2cm}|p{0.4cm}|p{0.65cm}|p{0.7cm}|p{0.7cm}|p{0.6cm}|p{0.5cm}|p{0.5cm}|p{0.6cm}|}\hline
\textit{\textbf{m}} & \textit{\textbf{digit size (n)}} & \textit{\textbf{total digits (d)}} & \textit{\textbf{Freq (MHz)}} & \textit{\textbf{latency ($\mu s$)}} & \textit{\textbf{LUTs}} & \textit{\textbf{Regs}} & \textit{\textbf{Carry}} & \textit{\textbf{Power (mW)}}\\\hline

\multirow{5}{*}{\rotatebox[origin=c]{90}{521$\times$521}} &  32 &  17 &  33.11 &  16.43 &  6369 &  1692 & \cellcolor{LightCyan!25} 408 &  184 \\ \cline{2-9}
{} & 41 & 13 & 29.15 & 18.28 & 7995 & 1681 & 416 & 192 \\ \cline{2-9}
{} & 53 & 10 & 28.32 & 22.72 & 8079 & 1732 & 417 & 191 \\ \cline{2-9}
{} &  64 &  9 &  34.48 &  15.12 &  6095 &  1758 &  408 &  220 \\ \cline{2-9}
{} & 81 & 8 & 30.30 & 21.38 & 8207 & 1795 & 415 & 247 \\ \cline{2-9}
{} &  128 &  5 &  34.84 &  14.95 &  5964 &  1881 &  424 &  220 \\ \hline

\multirow{4}{*}{\rotatebox[origin=c]{90}{571$\times$571}} & 32 & 17 & 30.12 & 18.06 & 6397 & 1847 & 447 & 194 \\ \cline{2-9}
{} & 41 & 13 & 27.17 & 19.62 & 8750 & 1834 & 455 & 192 \\ \cline{2-9}
{} & 53 & 10 & 26.04 & 20.35 & 9053 & 1880 & 449 & 187 \\ \cline{2-9}
{} & 81 & 8 & 28.01 & 23.13 & 8958 & 1951 & 452 & 226 \\ \hline

\multirow{10}{*}{\rotatebox[origin=c]{90}{1024$\times$1024}} & 2 & 512 & 14.22 & 72.11 & 10993 & 3634 & 1085 & 173 \\ \cline{2-9}
{} & 4 & 256 & 15.89 & 64.48 & 10824 & 3384 & 928  & 172 \\ \cline{2-9}
{} & 8 & 128 & 16.86 & 60.66 & 11074 & 3261 & 849 & 180 \\ \cline{2-9}
{} & 16 & 64 & 17.51 & 58.48 & 10634 & 3248 & 811 & 185 \\ \cline{2-9}
{} & 32 & 32 & 17.89 & 57.28 & 11371 & 3267 & 791 & 190 \\ \cline{2-9}
{} & 64 & 16 & 17.95 & 57.04 & 11947 & 3330 & 792 &195 \\ \cline{2-9}
{} & 128 & 8 & 18.57 & 55.14 & 12207 & 3450 & 800 & 221 \\ \cline{2-9}
{} & 256 & 4 & 18.93 & 54.09 & 11367 & 3740 & 832 & 247 \\ \cline{2-9}
{} & 512 & 2 & 19.12 & 53.55 & 10385 & 4295 & 896 & 226\\ \cline{2-9}
{} & 1024 & 1 & 18.46 & 55.50 & 11462 & 5303 & 1024 & 235 \\ \hline
\end{tabular}}
\end{table}

\subsubsection{ASIC non-digitized multipliers} \label{subsec:comparison_non_digitized}
Our results consider NIST-defined prime (P-192 to P-521) and binary (B-163 to B-571) fields utilized in ECC-based public key cryptosystems. As the operand sizes increase, the corresponding clock frequency decreases, as shown in column three of Table \ref{tab:table_1}. The decrease in frequency leads to an increase in latency, as presented in column four of Table \ref{tab:table_1}. In addition to latency, the corresponding area and power values also increase with the increase in size of multiplier operands (see columns five and six of Table \ref{tab:table_1}). It is evident from these results that the SBM multiplier requires less hardware resources than 2-way Karatsuba, 3-way Toom-Cook, and 4-way Toom-Cook multipliers. Moreover, the 2-way Karatsuba achieves lower power values as compared to other selected multipliers. This is explained by the datapath and the composition of the different multipliers. SBM requires $2m + 2m$ bit adder, 2-way Karatsuba requires $m + m + m$ bit adder/subtracter for generating final polynomial, 3-way Toom-Cook requires fifteen $\frac{m}{3}$ bit incrementers, and 4-way Toom-Cook requires sixteen $\frac{m}{4}$ bit incrementers. There is always a trade-off between various design parameters such as area, latency, power etc. Consequently, the SBM multiplier is more useful for area constrained applications. For better latency, other multipliers are more convenient.

\subsubsection{ASIC digitized multipliers}  \label{subsec:comparison_sbm_digitized}
For digitizing, we have selected 521, 571, and 1024 as the lengths of the input operands, as shown in column one of Table \ref{tab:table_2}. Moreover, for input lengths of 521 and 571, digit sizes of 32, 41, 53 and 81 have been adopted. For an input length of 1024 bits, digit sizes are given in powers of two, for $n$ = $2, \ldots, 1024$. Digit size $n$ and total digits $d$ are listed in columns two and three of Table \ref{tab:table_2}, respectively. It is noteworthy that the increase in digit size results in a decrease in clock frequency, as presented in column four of Table \ref{tab:table_2}. Moreover, it also translates to an increase in latency, as shown in column five of Table \ref{tab:table_2}. For the $1024\times1024$ multiplier, the obtained values for area and power show behavior similar to a parabolic curve with respect to digit size, as given in the last two columns of Table \ref{tab:table_2}. This is intuitive, as in the extreme cases of too small or too large digits, the wrapper logic becomes inefficient and may even become the bottleneck for timing. In summary, for an application that requires high clock frequency, shorter digits are preferred; however, this brings a significant cost in area and power. 

\subsubsection{FPGA digitized multipliers}  \label{subsec:comparison_sbm_digitized_fpga}

Alike ASIC demonstrations (presented in Sec. \ref{subsec:comparison_sbm_digitized}), we have chosen similar lengths of the input operands (521, 571, and 1024) for the evaluation on an Artix-7 FPGA platform, as shown in column one of Table \ref{tab:table_3}.We have used Xilinx Viviado Desig Suite for the FPGA based experiments.  Furthermore, for input lengths of 521 and 571, digit sizes of 32, 41, 53 and 81 have been considered. For an input length of 1024 bits, digit sizes are adopted in powers of two, for $n$ = $2, \ldots, 1024$. Digit size $n$ and total digits $d$ are listed in columns two and three of Table \ref{tab:table_3}, respectively. The synthesis results (clock frequency, latency, area in terms of LUTs and Regs, and power) achieved for FPGA are totally distinct when compared to ASIC values as the implementation platforms are quite contrasting. It is important to note that the frequency of the multiplier architecture increases with the increase in digit size (shown in column four of Table \ref{tab:table_3}). This phenomenon keeps on-going until it reaches a saturation point (i.e., best possible performance in terms of clock frequency with respect to $n$). Once it reaches a saturation point, then there is a decrease in the clock frequency. Moreover, the saturation occurs at any digit size between 0 to $n$ (in this work and for this experiment, the saturation occurs when the value for $n$ = $512$). The saturation point also varies with the change in operand size of the multiplier as given in Table \ref{tab:table_3}. For other reported parameters, i.e., latency, LUTs and power, the saturation point is not possible to show as there is a non-linear behavior (see columns five, six and nine of Table \ref{tab:table_3}). It is noteworthy that we have considered the worst case scenario by excluding the DSP (Digital Signal Processing) blocks during synthesis. The performance of multiplier architectures will be higher by considering the conventional synthesis flow with DSPs.

\subsubsection{Figure-of-Merit (FoM) for digitized SBM multiplier}  \label{subsec:fom} 
A FoM is defined to perform a comparison while taking into account different design characteristics at the same time. A FoM to evaluate the latency and area parameters for both ASIC and FPGA platforms is defined using Eq. \ref{eq:latency_times_area}. The higher the FoM values, the better. Similarly, a ratio for latency and power characteristics are calculated considering Eq. \ref{eq:latency_times_power}. 

        \begin{equation}\label{eq:latency_times_area}
            FoM = \frac{1}{latency\,(\mu s) \times area}
        \end{equation}

    
        \begin{equation}\label{eq:latency_times_power}
            FoM = \frac{1}{latency\,(\mu s) \times power\,(mW)}
        \end{equation}

\begin{figure}[ht]
\centering \footnotesize
\includegraphics[width=2.75in]{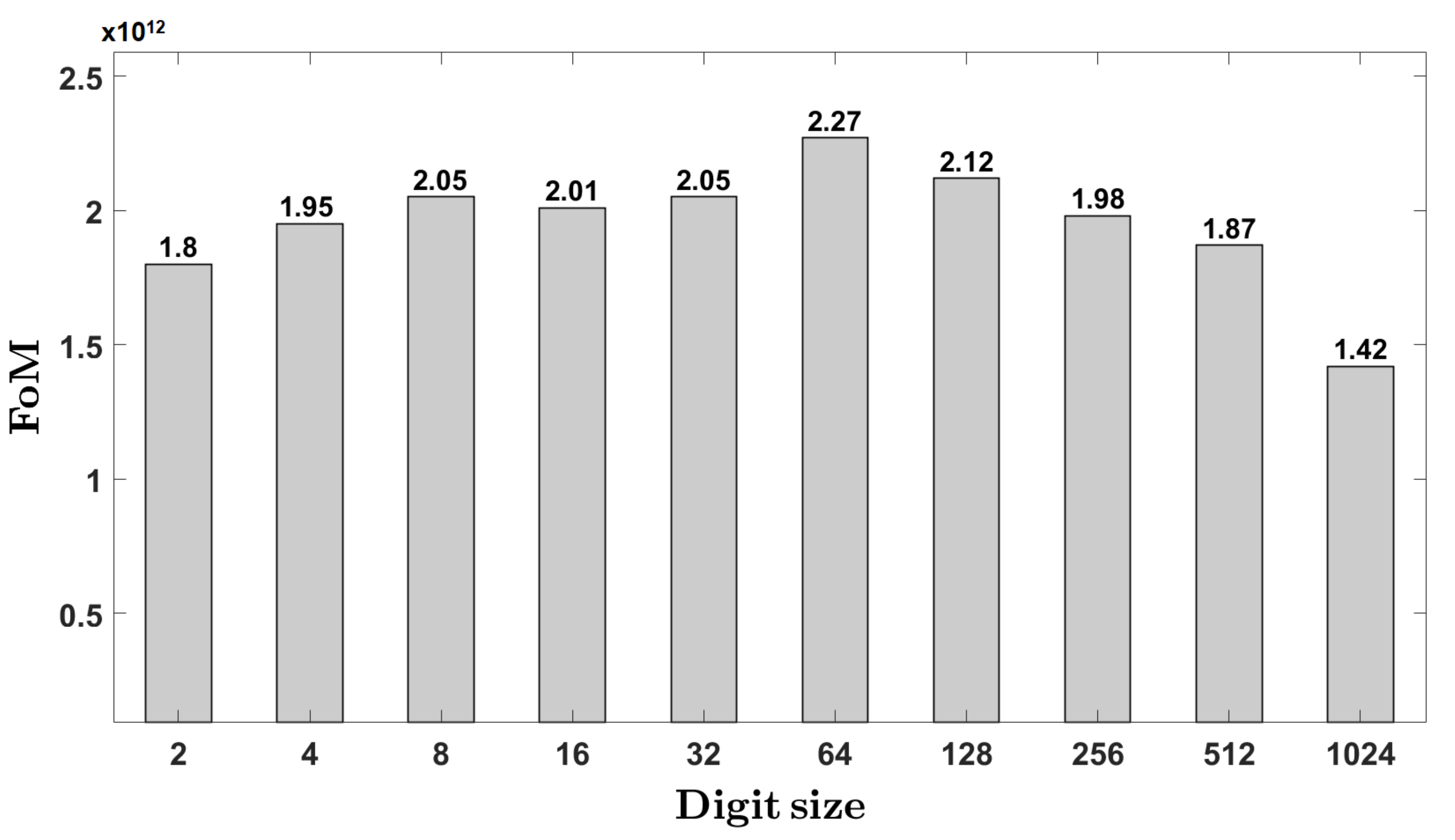}
\caption{Area and latency FoM for various digit sizes of a $1024\times 1024$ multiplier}
\centering
\label{fig:figure_2}
\end{figure}

\begin{figure}[ht]
\centering \footnotesize
\includegraphics[width=2.75in]{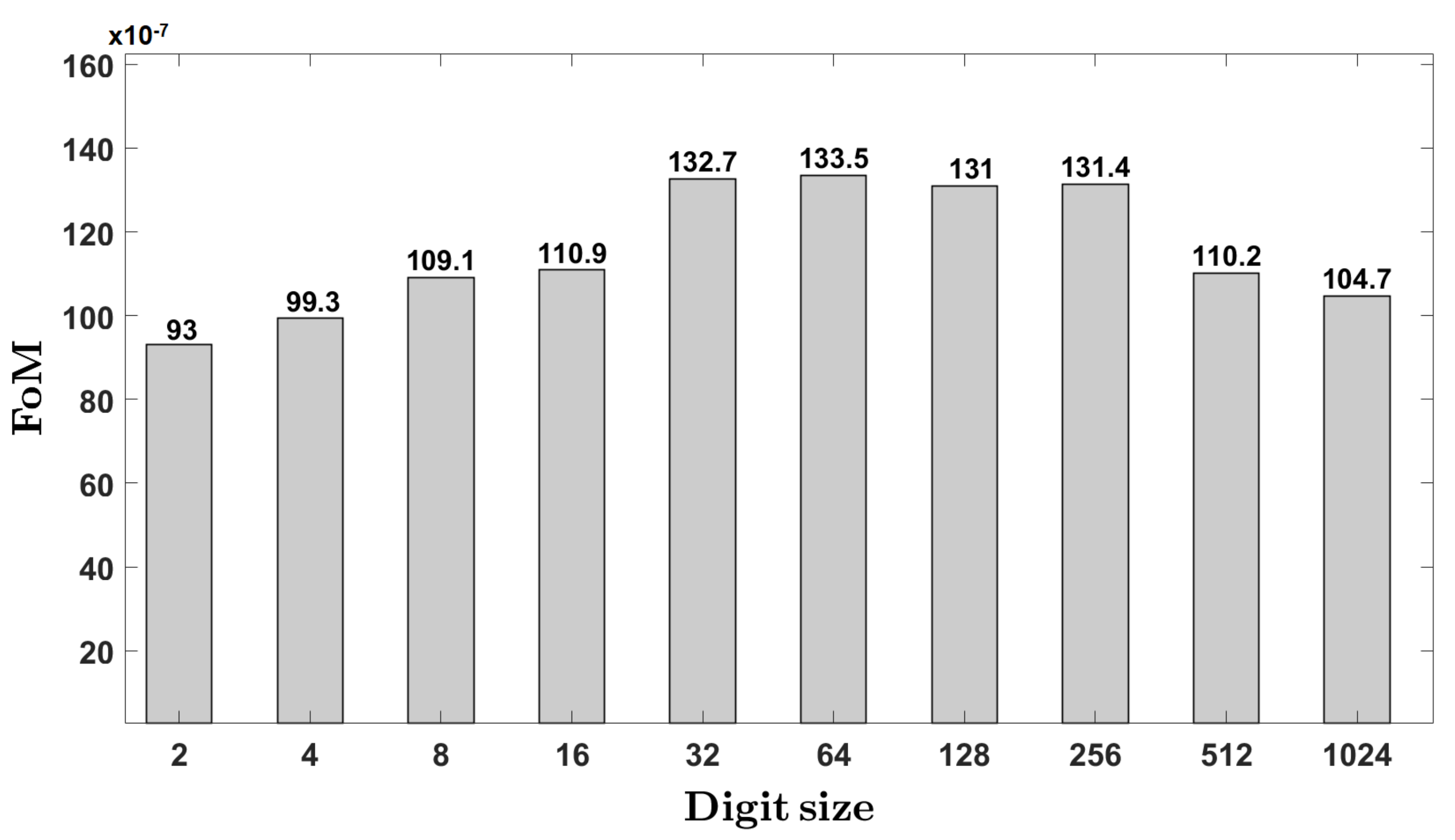}
\caption{Power and latency FoM for various digit sizes of a $1024\times 1024$ multiplier}
\centering
\label{fig:figure_3}
\end{figure}

 The calculated values of defined FoMs for ASIC are given in figures \ref{fig:figure_2} and \ref{fig:figure_3}, where various digit sizes were considered for a $1024\times1024$ multiplier. 
 
For both FoMs (shown in figures \ref{fig:figure_2} and \ref{fig:figure_3}), it becomes clear that the extreme cases lead to suboptimal results. For the studied 1024 $\times$ 1024 multiplier, the variant with $n=64$ and $d=16$ presents an optimal solution. Other similar values, such as $n=32$ and $n=128$, also give very close to optimal solutions.

\begin{figure}[]
\centering \footnotesize
\includegraphics[width=2.75in]{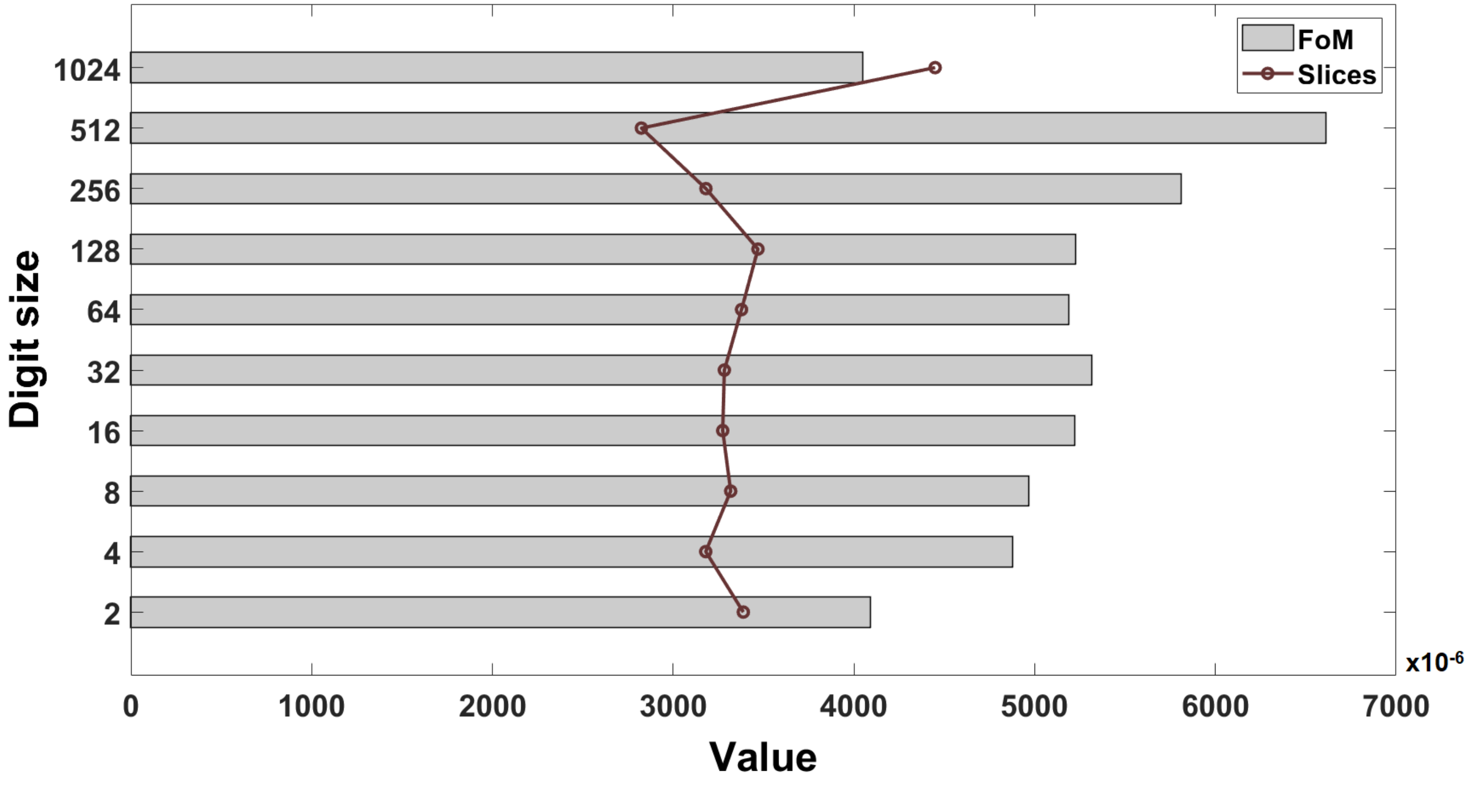}
\caption{Slices and latency FoM for various digit sizes of a $1024\times 1024$ multiplier}
\centering
\label{fig:figure_4}
\end{figure}

Likewise ASICs, the calculated values of defined FoM (from Eq. \ref{eq:latency_times_area}) for FPGA is given in Fig. \ref{fig:figure_4}, where various digit sizes were considered for a 1024$\times$1024 multiplier. To calculate FPGA area utilizations, the slices flip-flops, LUTs and carry units are the basic building-blocks. Therefore, the FoM in Eq. \ref{eq:latency_times_area} can be calculated by employing different metrics-of-interest (e.g., slices, LUTs, registers and carry blocks). Note that we have used an FPGA slices as area in Eq. \ref{eq:latency_times_area}. Fig. \ref{fig:figure_4} reveals that the FoM value for $n=512$ and $d=2$ results an optimal solution. 

\begin{figure}[t]
\centering \footnotesize
\includegraphics[width=2.75in]{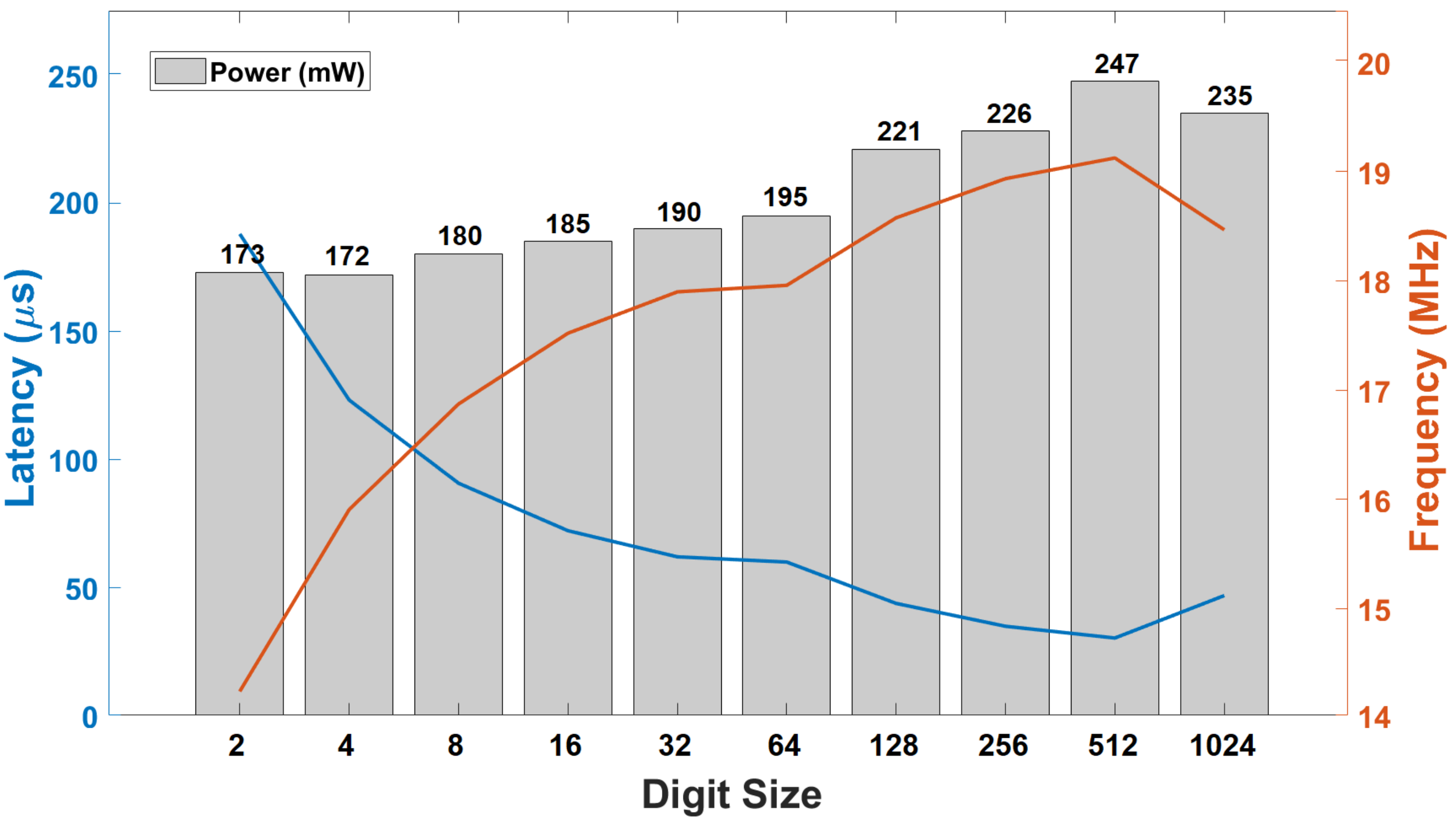}
\caption{Frequency, latency and power analysis for various digit sizes of a $1024\times 1024$ multiplier}
\centering
\label{fig:figure_6}
\end{figure}

The combined relation between frequency, latency and power for different values of $n$ is illustrated in Fig. \ref{fig:figure_6}. Therefore, it is noted from Fig. \ref{fig:figure_6} that the value of latency decreases, frequency increases with the increase in $n$. The increase in frequency and decrease in latency keeps on-going until saturation point occurs (when $n=512$). 

\subsection{Comparison to the state of the art}  \label{subsec:comparisons} 
To perform a fair comparison with existing state-of-the-art modular multiplier architectures, we have used similar operand lengths, digit sizes and implementation technologies (for FPGA and ASIC) as used in the corresponding solutions, shown in Table \ref{tab:table_4}. In state-of-the-art solutions, multiplication results are given for different operands length. However, we have provided comparison of our results with only the larger operands. Moreover, we have used symbol `N/A' in Table \ref{tab:table_4} where the values for design parameters (\textit{Freq}, \textit{latency} and \textit{area}) are not given.

\begin{table}[ht]
\begin{threeparttable}
\caption{Area and latency comparisons of non-digitized and digitized multipliers with state of the art}
{\begin{tabular}{|p{0.38cm}|p{1.68cm}|p{0.7cm}|p{0.4cm}|p{0.6cm}|p{0.7cm}|p{1.57cm}|}\hline
\textit{\textbf{Ref}} & \textit{\textbf{Multiplier}} & \textit{\textbf{Device}} & \textit{\textbf{m}} & \textit{\textbf{Freq (MHz)}} & \textit{\textbf{latency ($\mu s$)}} & \textit{\textbf{Area ($\mu m^2$)/LUTs}}\\\hline

\multirow{3}{*}{\cite{Rafferty_2017}} & 2-way KM & V7 & {128} & 104.3 & 0.61 & 3499 \\ \cline{2-7}
{} & {2-way KM} & {V7} & {256} & 74.5 & 1.71 & 7452 \\ \cline{2-7}
{} & {2-way KM} & {V7} & {512} & 51.6 & 4.96 & 20474 \\ \hline


\multirow{1}{*}{\cite{ASIC_65nm_2014}} & BL-PIPO & 65nm & 163 & {N/A} & {N/A} & {5328 GE} \\\hline

\multirow{1}{*}{\cite{Morales_Sandoval}} & DSM (ds=64) & V6 & 571 & {258.5} & {0.03} & {10983} \\\hline

\multirow{3}{*}{\cite{Pan}} & DSMM (ds=2) & V7 & 2048 & {N/A} & {N/A} & {18067} \\\cline{2-7}
{} & DSMM (ds=4) & V7 & 2048 & {N/A} & {N/A} & {33734} \\\cline{2-7}
{} & DSMM (ds=8) & V7 & 2048 & {N/A} & {N/A} & {62023} \\\hline
\multirow{8}{*}{\textbf{TW}} & SBM & 65nm & 163 & {N/A} & {N/A} & {11727 GE} \\\cline{2-7}

{} & {2-way KM} & {V7} & {128} & 167.4 & 0.38 & 2110 \\ \cline{2-7}
{} & {2-way KM} & {V7} & {256} & 119.9 & 1.06 & 4318 \\ \cline{2-7}
{} & {2-way KM} & {V7} & {512} & 63.8 & 4.01 & 9582 \\ \cline{2-7}


{} & SBM (ds=2) & V7 & 2048 & {15.03} & {69760} & {25559} \\\cline{2-7}
{} & SBM (ds=4) & V7 & 2048 & {16.6} & {15790} & {22040} \\\cline{2-7}
{} & SBM (ds=8) & V7 & 2048 & {17.4} & {3760} & {23315} \\\cline{2-7}

{} & SBM (ds=64) & V6 & 571 & {46.4} & {1.74} & {6181} \\\cline{1-7}

\end{tabular}}
\label{tab:table_4}

\begin{tablenotes}
      \small
      \item \textbf{V7:} Xilinx Virtex-7, \textbf{V6:} Xilinx Virtex-6, \textbf{ds:} digit size, \textbf{TW:} this work, \textbf{DSM:} Digit Serial Montgomery multiplier based wrapper, \textbf{BL-PIPO:} Bit level parallel in parallel out multiplier using SBM multiplication method, \textbf{GE:} gate equivalents
    \end{tablenotes}
  \end{threeparttable}
\end{table}

Concerning only the non-digitized multipliers for comparison, the 2-way Karatsuba multiplier of \cite{Rafferty_2017} over Virtex-7 FPGA for operand sizes of 128, 256 and 512 bits presents 38\%, 39\% and 20\% higher latency when compared to 2-way Karatsuba multiplier generated by TTech-LIB, as shown in Table \ref{tab:table_4}. Moreover, the generated multiplier utilizes lower hardware resources in terms of LUTs (see column seven in Table \ref{tab:table_4}) as compared to resources (LUTs) utilized in \cite{Rafferty_2017}. 
On 65nm node, the BL-PIPO multiplier of \cite{ASIC_65nm_2014} utilizes 55\% lower hardware resources in terms of gate counts as compared to our SBM multiplier generated by TTech-LIB. 

When digitized flavor of polynomials multiplication is considered for comparison over different digit sizes, the digit serial Montgomery multiplier based wrapper of \cite{Morales_Sandoval} results 83\% higher clock frequency and requires 58\% less computational time as compared to our SBM based digit serial wrapper generated by TTech-LIB. On the other hand, the SBM based digit serial wrapper results 56\% lower hardware resources over Virtex-6 FPGA. There is always a trade-off between performance and area parameters. Another digit serial modular multiplication based wrapper of \cite{Pan} results 14\% (for ds=2) lower FPGA LUTs while for remaining digit sizes of 4 and 8, it utilizes 35\% and 63\% higher FPGA LUTs as compared to SBM wrapper generated by TTech-LIB. The frequency and latency parameters cannot be compared as these are not given. 

The comparisons and discussion above show that the multipliers generated by TTech-LIB provide a realistic and reasonable comparison to state-of-the-art multiplier solutions \cite{Rafferty_2017,ASIC_65nm_2014,Morales_Sandoval,Pan}. Hence, not only can users explore various design parameters within our library, they can also benefit from implementations that are competitive with respect to the existing literature.

\section{Conclusion} \label{conclusion}
This work has presented an open-source library for large integer polynomial multipliers. The library contains digitized and non-digitized flavors of polynomial coefficient multipliers. For non-digitized multipliers, based on the values for various design parameters, users/designers can select amongst several studied multipliers according to needs of their targeted application. Furthermore, we have shown that for digitized multipliers, the evaluation of individual design parameters may not be comprehensive, and figures of merit are better suited to capture the characteristics of a circuit. Furthermore, we believe the results enabled by TTech-LIB will guide hardware designers to select an appropriate digit size that reaches an acceptable performance according to application requirements. This is achieved with the aid of TTech-LIB's generator, which helps a designer to quickly explore the complex design space of polynomial multipliers.


\bibliographystyle{IEEEtran}
\bibliography{multiplier}

\begin{thebibliography}{10}
\providecommand{\url}[1]{#1}
\csname url@samestyle\endcsname
\providecommand{\newblock}{\relax}
\providecommand{\bibinfo}[2]{#2}
\providecommand{\BIBentrySTDinterwordspacing}{\spaceskip=0pt\relax}
\providecommand{\BIBentryALTinterwordstretchfactor}{4}
\providecommand{\BIBentryALTinterwordspacing}{\spaceskip=\fontdimen2\font plus
\BIBentryALTinterwordstretchfactor\fontdimen3\font minus
  \fontdimen4\font\relax}
\providecommand{\BIBforeignlanguage}[2]{{%
\expandafter\ifx\csname l@#1\endcsname\relax
\typeout{** WARNING: IEEEtran.bst: No hyphenation pattern has been}%
\typeout{** loaded for the language `#1'. Using the pattern for}%
\typeout{** the default language instead.}%
\else
\language=\csname l@#1\endcsname
\fi
#2}}
\providecommand{\BIBdecl}{\relax}
\BIBdecl

\bibitem{RSA_ECC}
H.~Eberle, N.~Gura, S.~Shantz, V.~Gupta, L.~Rarick, and S.~Sundaram, ``A
  public-key cryptographic processor for rsa and ecc.''\hskip 1em plus 0.5em
  minus 0.4em\relax IEEE, 2004, pp. 98--110.

\bibitem{NIST_Competition}
\BIBentryALTinterwordspacing
NIST, ``Computer security resource centre: Pqc standardization process, third
  round candidate announcement,'' 2020. [Online]. Available:
  \url{https://csrc.nist.gov/news/2020/pqc-third-round-candidate-announcement}
\BIBentrySTDinterwordspacing

\bibitem{cloud-computations}
A.~L\'{o}pez-Alt, E.~Tromer, and V.~Vaikuntanathan, ``On-the-fly multiparty
  computation on the cloud via multikey fully homomorphic encryption,'' in
  \emph{Proceedings of the Forty-Fourth Annual ACM Symposium on Theory of
  Computing}, ser. STOC '12.\hskip 1em plus 0.5em minus 0.4em\relax New York,
  NY, USA: Association for Computing Machinery, 2012, p. 1219–1234.

\bibitem{NIST_Competition_Round_2}
\BIBentryALTinterwordspacing
NIST, ``Computer security resource centre: Post-quantum cryptography, round 2
  submissions,'' 2020. [Online]. Available:
  \url{https://csrc.nist.gov/projects/post-quantum-cryptography/round-2-submissions}
\BIBentrySTDinterwordspacing

\bibitem{Rafferty_2017}
C.~{Rafferty}, M.~{O’Neill}, and N.~{Hanley}, ``Evaluation of large integer
  multiplication methods on hardware,'' \emph{IEEE Transactions on Computers},
  vol.~66, no.~8, pp. 1369--1382, 2017.

\bibitem{2_way_Karatsuba}
M.~Imran, Z.~U. Abideen, and S.~Pagliarini, ``An experimental study of building
  blocks of lattice-based nist post-quantum cryptographic algorithms,''
  \emph{Electronics}, vol.~9, no.~11, p. 1953, Nov 2020.

\bibitem{Montgomery}
A.~A. Abd-Elkader, M.~Rashdan, E.-S.~A. Hasaneen, and H.~F. Hamed, ``Advanced
  implementation of montgomery modular multiplier,'' \emph{Microelectronics
  Journal}, vol. 106, p. 104927, 2020.

\bibitem{NTT}
A.~C. Mert, E.~Öztürk, and E.~Savaş, ``\uppercase{FPGA} implementation of a
  run-time configurable ntt-based polynomial multiplication hardware,''
  \emph{Microprocessors and Microsystems}, vol.~78, p. 103219, 2020.

\bibitem{ASIC_65nm_2014}
R.~{Azarderakhsh}, K.~U. {Järvinen}, and M.~{Mozaffari-Kermani}, ``Efficient
  algorithm and architecture for elliptic curve cryptography for extremely
  constrained secure applications,'' \emph{IEEE Transactions on Circuits and
  Systems I: Regular Papers}, vol.~61, no.~4, pp. 1144--1155, 2014.

\bibitem{Interleaved_modular_reduction_algorithm}
\BIBentryALTinterwordspacing
S.~R. Pillutla and L.~Boppana, ``An area-efficient bit-serial sequential
  polynomial basis finite field gf(2m) multiplier,'' \emph{AEU - International
  Journal of Electronics and Communications}, vol. 114, p. 153017, 2020.
  [Online]. Available:
  \url{http://www.sciencedirect.com/science/article/pii/S1434841119318485}
\BIBentrySTDinterwordspacing

\bibitem{homomorphic_encryption}
Y.~{Doröz}, E.~{Öztürk}, and B.~{Sunar}, ``Accelerating fully homomorphic
  encryption in hardware,'' \emph{IEEE Transactions on Computers}, vol.~64,
  no.~6, pp. 1509--1521, 2015.

\bibitem{Xie_2018}
J.~{Xie}, P.~K. {Meher}, X.~{Zhou}, and C.~{Lee}, ``Low register-complexity
  systolic digit-serial multiplier over $gf(2^m)$ based on trinomials,''
  \emph{IEEE Transactions on Multi-Scale Computing Systems}, vol.~4, no.~4, pp.
  773--783, 2018.

\bibitem{Morales_Sandoval}
M.~Morales-Sandoval, C.~Feregrino-Uribe, P.~Kitsos, and R.~Cumplido,
  ``Area/performance trade-off analysis of an fpga digit-serial gf(2m)
  montgomery multiplier based on lfsr,'' \emph{Computers \& Electrical
  Engineering}, vol.~39, no.~2, pp. 542 -- 549, 2013.

\bibitem{Pan}
J.~{Pan}, P.~{Song}, and C.~{Yang}, ``Efficient digit-serial modular
  multiplication algorithm on fpga,'' \emph{IET Circuits, Devices Systems},
  vol.~12, no.~5, pp. 662--668, 2018.

\bibitem{TTech_LIB}
\BIBentryALTinterwordspacing
M.~Imran, Z.~U. Abideen, and S.~Pagliarini,
  ``T\uppercase{T}ech-\uppercase{LIB}: Center for hardware security,'' 2020.
  [Online]. Available:
  \url{https://github.com/Centre-for-Hardware-Security/TTech-LIB}
\BIBentrySTDinterwordspacing

\bibitem{NIST_ECC_PARAMETERS}
\BIBentryALTinterwordspacing
C.~Lily, M.~Dustin, R.~Andrew, and R.~Karen, ``Recommendations for discrete
  logarithm-based cryptography: Elliptic curve domain parameters,'' 2020.
  [Online]. Available:
  \url{https://nvlpubs.nist.gov/nistpubs/SpecialPublications/NIST.SP.800-186-draft.pdf}
\BIBentrySTDinterwordspacing

\end{thebibliography}
\end{document}